# BOSE-EINSTEIN CONDENSATION IN CLASSICAL SYSTEMS

K.Staliunas
Physikalisch Technische Bundesanstalt, 38116 Braunschweig, Germany
tel.: +49-531-5924482, Fax: +49-531-5924423, E-mail: Kestutis.Staliunas@PTB.DE

**Abstract**

It is shown, that Bose-Einstein statistical distributions can occur not only in *quantum* system, but in *classical* systems as well. The coherent dynamics of the system, or equivalently autocatalytic dynamics in momentum space of the system is the main reason for the Bose-Einstein condensation. A coherence is possible in both quantum and classical systems, and in both cases can lead to Bose-Einstein statistical distribution.

Under Bose-Einstein Condensate (BEC) an *equilibrium* state of a *quantum* system is usually understood, such that most of the particles of the system condense into one, the *lowest* energy state of the system [1].

In the accompanying papers [2,3] we show that thermal equilibrium is not a necessary condition for Bose-Einstein condensation: the condensate can as well occur in a system far from thermal equilibrium. Additionally we show, that the condensation far from thermal equilibrium can occur not necessarily into the lowest state of the system, but also into some higher, excited state. This first paper from the series focuses on the assumption on the quantum character of the system for BEC.

A quantum character of the system is usually assumed to be an essential ingredient for the BEC. The indistinguishability of the particles in the same quantum state is the central assumption used in deriving the Bose-Einstein distribution originally [1]. We show in this article, that the quantum character in not a necessary assumption for obtaining Bose-Einstein distributions. Bose-Einstein- or Bose-Einstein-like distributions may be obtained in classical coherent (or ordered) systems as well.

Under classical coherent system we understand that the individual particles arrange into ordered structures in the space domain. Ordered structures are characterized not only by the amplitude of the order parameter (corresponding to the density of particles) but also by the phase of the order parameter (due to break-up of translation invariance). Such ordered structures correspond to sharp distributions in the momentum domain (the domain of the spatial wave-vectors). Therefore if in a self-organizing system the particles tend to arrange into ordered ensembles, then equivalently in the momentum domain the occupations tend to autocatalyze within the same momentum state.

This motivates the search for sharp (Bose-Einstein-like) distributions in classical coherent systems. Concretely, we demonstrate in the article, that the essential condition for a Bose-Einstein statistical distribution is that the random migration of the particles through the momentum states of the (quantum or classical) system is dependent on the occupation degree of these states. In most disordered (noncoherent) systems all particles can randomly migrate through the states with the migration rates independent on the occupation of the states. This is actually typical for linear systems, and this results in Boltzman statistical distributions. If, however, the particles are scattered from a particular state with the probabilities decreasing with the increasing number of particles in that state, then this can result in Bose-Einstein- or Bose-Einstein-like distribution. (As Bose-Einstein-like distributions we understand such distributions, for which most of the particles collect in one state.) Therefore, if the rate of the random migration of the particles through the states is dependent on the occupation of the states, which is typical for many nonlinear systems, then the resulting distributions can become Bose-Einstein-like.

A monotonically decreasing dependence of the scattering probability on the occupation number of a state is possible both in quantum and in classical systems. In the quantum case the collision cross-section of the bosons decreases with increasing occupation number of the state. In this way the quantum particles are randomly scattered at a lower rate from a state with high occupation, thus tend to collect, or to condense in the highly occupied states. In the classical case, members of the same (momentum) state autocatalyze the occupation of the state by additional members in the coherent case. As a result the probability to lose a member from a group decreases with increasing number of members of the group. This, as shown below, also leads to collection of the members in one group, thus to Bose-Einstein-like statistical distributions.

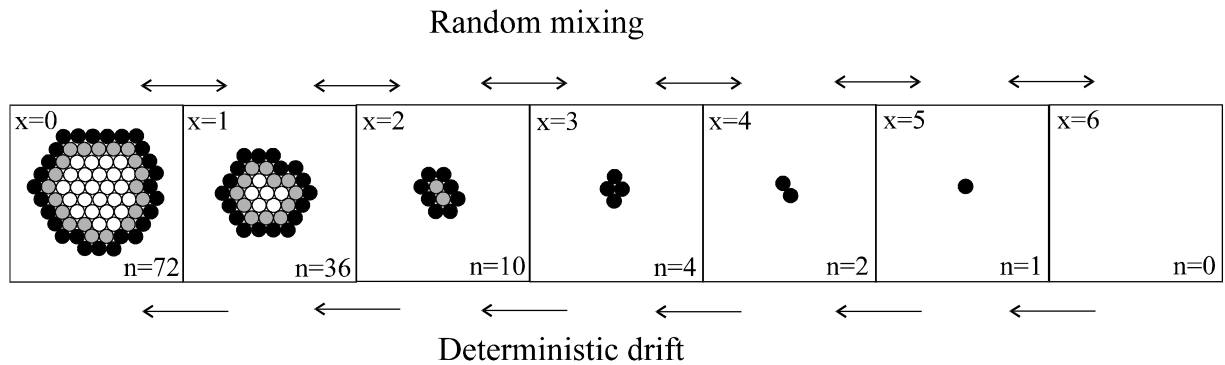

**Fig.1**: Random migration of marbles among boxes (momentum states). If the probability for a marble to be moved randomly to the neighboring box does not depends on the number of the marbles in the box, then a Boltzman distribution results. If the probability for random migration decreases with the increasing number of marbles in a box, e.g. assuming that only the marbles at the boundary (the black, and partially the gray ones) are randomly mixed, and the bulk marbles (the white ones) are screened from the random mixing (due to some classical autocatalytic interaction), then the distribution can become Bose-Einstein-like.

Here we assume that in general the probability for a particular member of a group (numbered by $x$) to be removed randomly from the particular group is dependent on the occupation of the group $n(x)$. $x$ numbers the discrete or continuous states in the discrete or continuous cases respectively. For clarity we assume first a random mixing of classical particles (marbles) among the boxes as shown in the Fig.1. In a discrete case, each individual particle can be randomly shifted to the left ($x-1$) or the right ($x+1$) box with a given constant probability. In addition to the random force, a deterministic drift is also assumed, caused e.g. by a potential, and moving the particles towards lower values of $x$. This is actually a model of random biased walk for each individual particle on a discrete one-dimensional lattice. In the continuous limit the process is described by a stochastic Langevin equation:

$$\frac{\partial x(t)}{\partial t} = -v + \Gamma(t) \qquad (1)$$

for the position of the particle $x(t)$. $v$ is the coefficient of the deterministic drift towards the smaller values of the coordinate $x$; and $\Gamma(t)$ is the $\delta$-correlated random force: $<\Gamma(t_1)\Gamma(t_2)> = q_0 \delta(t_1 - t_2)$, where the proportionality coefficient $q_0$ corresponds to the intensity (or temperature) of the random force. This model (1) is also often used for the random Brownian motion of a particle (e.g. in [4]), where $x(t)$ stands for the velocity of the randomly kicked Brownian particle, $q_0$ is the amplitude of the random kicks (proportional to the temperature of the environment). The deterministic drift in (1) towards the lower values of $x$ correspond to the viscosity for the Brownian particle.

The statistical distribution of the particles $n(x,t)$ then obeys:

$$\frac{\partial n(x,t)}{\partial t} = \left[\frac{\partial}{\partial x}v + \frac{\partial^2}{\partial x^2}q_0\right]n(x,t) \qquad (2)$$

(2) is the Fokker-Planck equation for (1) (see e.g. [4]). The stationary solution of (2) under the assumption for asymptotic: $n(x,t) \to 0$, and $\partial n(x,t)/\partial x \to 0$ for $x \to \infty$, is straightforward:

$$n(x) = n_0 \exp[-vx/q_0] \qquad (3)$$

which is the well known Boltzman distribution. $n_0$ is the occupation density (or occupation number in the discrete case) of the lowest state ($x \to 0$), and is related with the total number of particles $N$ by: $n_0 = Nv/q_0$.

Next we assume that the random migration of the particles through the states is dependent on the occupation of these states. We assume that the intensity of the random force acting on the particle depends on the occupation of the state $n(x,t)$ in the following way: $q(n) \propto q_0 \ln(n+1)/n$. (The motivation for this particular dependence is given below.) Then the correlation of the random forces in (1) becomes $n$-dependent: $<\Gamma(t_1)\Gamma(t_2)> = q(n)\delta(t_1 - t_2)$. This leads to the corresponding Fokker-Planck equation:

$$\frac{\partial n(x,t)}{\partial t} = \left[\frac{\partial}{\partial x}v + \frac{\partial^2}{\partial x^2}\frac{q_0 \ln(n(x,t)+1)}{n(x,t)}\right]n(x,t). \qquad (4)$$

The drift coefficient in (4) is calculated using the Ito definition [4]. The stationary solution of (4) under the asymptotics: $n(x,t) \to 0$, and $\partial n(x,t)/\partial x \to 0$ for $x \to \infty$ is:

$$n(x) = \frac{1}{\alpha \cdot \exp[vx/q_0] - 1} \qquad (5)$$

which is identical to the well known Bose-Einstein distribution originally derived for quantum systems [1]. The parameter $\alpha$ depends on the total number of the particles: $\alpha = (1 - \exp[-Nv/q_0])^{-1}$ and is related with the chemical potential $\mu$ by $\alpha = \exp[\mu]$. The occupation of the lowest state is: $n_0 = (\exp[Nv/q_0] - 1)$. (5) coincides with the Boltzman distribution (3) in the limit of $Nv/q_0 \ll 1$, and shows a shows sharp increase of the occupation number of the lowest state $n_0$ in the opposite limit $Nv/q_0 \gg 1$. In this way, the condensation threshold for (4) can be defined as $Nv/q_0 \approx 1$.

In this way, if the scattering probability monotonically decreases with the number of the particles in a state, then the statistical distributions can become of Bose-Einstein type. This decrease of scattering probability can occur in classical systems (due to coherence in a classical sense), as well as in quantum systems (due to indistinguishability of bosons).

The particular dependence of the scattering probability $q(n) = q_0(\ln(n+1))/n$, used above, generates the exact form of the Bose-Einstein distribution. This particular dependence means that the scattering probability is constant in the limit of weak occupations $n \ll 1$, and proportional to $(\ln(n)/n)^{1/2}$) in the limit of high occupations $n \gg 1$. This particular choice for the form of the scattering probability in the classical case can be understood by a gedanken experiment illustrated in Fig.1. Here it is assumed that the marbles are not completely randomly mixed among the boxes, but that to some degree they autocatalyze within each box. This autocatalytic action may occur due to some attractive interaction binding the marbles into clusters. Also, by assuming that only the marbles at the boundary are subjected to random scattering, screening the bulk marbles from the destructive influence of the environment.

If the above used dependence $q(n) = q_0 (\ln(n+1))/n$ generates the exact form of Bose-Einstein distributions, then every other form of scattering probability decreasing monotonically with increasing occupation number $n$ can lead to distribution similar to the Bose-Einstein distributions. For a general form of the diffusion coefficient $q(n)$ the Fokker-Planck equation becomes:

$$\frac{\partial n(x,t)}{\partial t} = \left[\frac{\partial}{\partial x} v + \frac{\partial^2}{\partial x^2} q(n)\right] n(x,t) \tag{6}$$

For a general form of $q(n)$ no explicit stationary solution of (6) exists. However, assuming that in the leading order (for large values of $n$) the form of the $q(n)$ is a power law: $q(n) \propto q_0 n^{-\gamma}$, simple expressions for the asymptotical distributions at $x \to 0$ are possible:

$$n(x) \propto \left[\frac{\gamma(1-\gamma)q_0}{vx}\right]^{\frac{1}{\gamma}} \tag{7}$$

This results in the Bose-Einstein-like distributions (with a singularity at $x \to 0$) for the values of $\gamma$ in the region: $0 < \gamma < 1$. The above studied case leading to exact Bose-Einstein distributions was indeed obtained for $\gamma = 1 - \varepsilon$. For $\gamma \geq 1$, when the autocatalytic action of the individuals within one group is too strong, no smooth statistical distributions can be obtained. The stationary solution of (6) becomes a $\delta$-function in this case: $n(x) \propto \delta(x)$, which means a condensation of all particles into the lowest state, independent on the temperature of the system.

Concluding: Bose-Einstein-like distributions can occur in coherent systems both in classical or quantum cases. Quantum mechanics is not necessary to obtain Bose-Einstein distributions. Essential is that due to a collective behavior members of each group autocatalyze occupation of their states and can not be easily removed from the group by stochastic forces. In the quantum case this "mutual support" appears through indistinguishability of the individual particles, reducing the scattering cross-section of the particles and thus reducing the scattering probability in highly occupied states.

In the accompanying papers [2,3] we show, that the Bose-Einstein-like distributions may occur not only in equilibrium classical systems as studied in the present article, but also in the systems far from thermal equilibrium.

We acknowledge discussions with C.O.Weiss and M.Lewenstein. This work has been supported by Sonderforschungs Bereich 407.